\documentclass{IEEEtran}
\IEEEoverridecommandlockouts
\usepackage{cite}
\usepackage{amsmath,amssymb,amsfonts}
\usepackage{algorithmic}
\usepackage{graphicx}
\usepackage{textcomp}
\usepackage{xcolor}
\usepackage{textcomp}
\usepackage{adjustbox}

\usepackage{booktabs}
\usepackage{multirow}
\def\BibTeX{{\rm B\kern-.05em{\sc i\kern-.025em b}\kern-.08em
    T\kern-.1667em\lower.7ex\hbox{E}\kern-.125emX}}

\begin{document}

\title{ExpGest: Expressive Speaker Generation Using Diffusion Model and Hybrid Audio-Text Guidance}
\author{
    \begin{minipage}{0.3\textwidth}
        \centering
        Yongkang Cheng\IEEEauthorrefmark{1} \\
        \textit{Northwest A\&F University} \\
        Shenzhen, China \\
        ykcheng@tencent.com
    \end{minipage}
    \hfill
    \begin{minipage}{0.3\textwidth}
        \centering
        Mingjiang Liang\IEEEauthorrefmark{1} \\
        \textit{University of Technology Sydney} \\
        Sydney, Australia \\
        mingjiang.liang@student.uts.edu.au
    \end{minipage}
    \hfill
    \begin{minipage}{0.3\textwidth}
        \centering
        Shaoli Huang \\
        \textit{Tencent AILab} \\
        Shenzhen, China \\
        shaol.huang@gmail.com
    \end{minipage}
    \hfill
    \begin{minipage}{0.3\textwidth}
        \centering
        Gaoge Han \\
        \textit{City University of Hong Kong} \\
        Hongkong, China \\
        hangaoge@gmail.com
    \end{minipage}
    \begin{minipage}{0.4\textwidth}
        \centering
        Jifeng Ning\textsuperscript{\textdagger} \\
        \textit{Northwest A\&F University} \\
        Xianyan, China \\
        njf@nwsuaf.edu.cn
    \end{minipage}
    \hfill
    \begin{minipage}{0.4\textwidth}
        \centering
        Wei Liu \\
        \textit{University of Technology Sydney} \\
        Sydney, Australia \\
        wei.liu@uts.edu.au
    \end{minipage}
    \thanks{\IEEEauthorrefmark{1}These authors contributed equally and \textsuperscript{\textdagger} is corresponding author.}
}

\maketitle

\begin{abstract}
Existing gesture generation methods primarily focus on upper body gestures based on audio features, neglecting speech content, emotion, and locomotion. These limitations result in stiff, mechanical gestures that fail to convey the true meaning of audio content. We introduce \textbf{ExpGest}, a novel framework leveraging synchronized text and audio information to generate expressive full-body gestures. Unlike AdaIN or one-hot encoding methods, we design a noise emotion classifier for optimizing adversarial direction noise, avoiding melody distortion and guiding results towards specified emotions.  Moreover, aligning semantic and gestures in the latent space provides better generalization capabilities. ExpGest, a diffusion model-based gesture generation framework, is the first attempt to offer mixed generation modes, including audio-driven gestures and text-shaped motion. Experiments show that our framework effectively learns from combined text-driven motion and audio-induced gesture datasets, and preliminary results demonstrate that ExpGest achieves more expressive, natural, and controllable global motion in speakers compared to state-of-the-art models. Our code, model, and demo are are available at https://github.com/cyk990422/ExpGest/.
\end{abstract}
%
%

\section{Introduction}
\label{sec:intro}
\vspace{-5pt}
In fields like virtual agents, movies, and human-computer interaction, virtual speakers convey information through co-speech gestures related to audio melody and content, and non-spontaneous movements. The emerging field of audio gesture generation has gained attention, with early research using rule-based methods~\cite{kipp2005gesture}, while data-driven techniques~\cite{habibie2022motion} improved diversity using statistical models. Deep models, such as VAE~\cite{kingma2013auto}, flow models~\cite{ye2022audio}, and diffusion-based models~\cite{yang2024freetalker, cheng2024conditional}, generate gestures directly from raw audio data. Methods combining audio melody and semantics~\cite{ao2022rhythmic} have advanced significantly. However, DiffStyleGesture~\cite{yang2023diffusestylegesture} and Emog~\cite{yin2023emog}, which use emotion as guidance, perform poorly on the BEAT dataset~\cite{liu2022beat}. 
\begin{figure}
  \begin{center}
  \vspace{-2em}
    \includegraphics[width=1\linewidth]{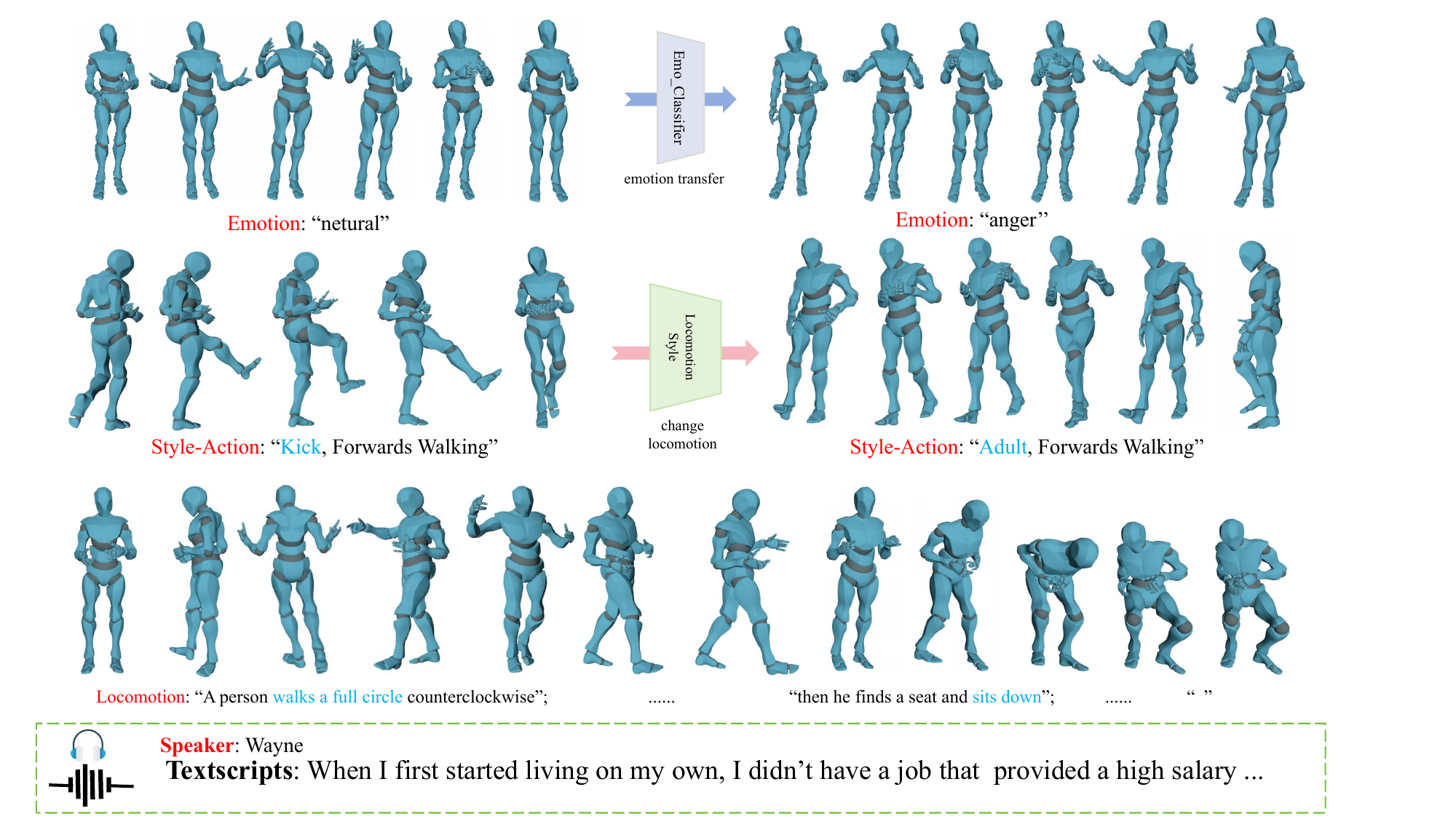}
    \vspace{-1em}
  \end{center}
  \vspace{-20pt}
  \caption{\textbf{Our method demonstrates multimodal-driven effects.} The first row showcases the emotional control capability of gestures driven by audio alone, while the second row exhibits the motion style transferability driven by a combination of phrases and audio. The third row presents the results of long-frame textual descriptions and audio jointly driving the process.}
  \vspace{-15pt}
  \label{fig:title_img}
  \end{figure}
\\In non-spontaneous motion generation, early methods~\cite{ghosh2017learning} considered deterministic signal-to-motion mapping using neural networks. The complexity and randomness of motion led to deeper research on generative models. TEMOS~\cite{petrovich2022temos} pioneered action-to-action and text-to-action tasks. Recently, KIT-ML~\cite{plappert2016kit} and HumanML3D~\cite{guo2022generating} datasets promoted text-guided motion generation. T2M-GPT~\cite{zhang2023t2m} uses autoregressive models, improving results but with jitter issues. Diffusion model-based methods~\cite{tevet2022human,chen2023executing,han2024hutumotion,han2025reindiffuse,cheng2025conditional} enhance motion data continuity through denoisers. Despite advancements, bottlenecks persist. No work has coherently integrated both motion categories, and different motion representations lead to dataset inconsistency, keeping the field in the single modality generation stage.\\
In this paper, we propose a diffusion model-based method aimed at using input text, audio, or a combination of both to guide the generation of expressive and diverse high-quality speakers.  As shown in Figure~\ref{fig:title_img}, when only audio is input, our model aims to generate highly expressive and rich gestures. When inputting a text-audio mixed modality, it generates co-speech gesture results with non-spontaneous motion. Specifically, for text-generated non-spontaneous motion, we adopt MDM as the baseline and address its lack of melody perception by encoding spectral features into text embeddings, thus alleviating the global motion issue in generated speakers unrelated to melody. For audio-generated gestures, we observe an objective fact: fingers and limbs exhibit different sensitivities to the two attributes (melody and semantics) in the audio. For instance, when calmly saying ``one, two, three", the arm tends to remain relatively still, while the fingers display the primary changes. In contrast, when the tone varies, significant changes occur in arm movements. Based on this observation, ExpGest is the first method to decouple fingers and limbs, assigning different weights to semantics and melody to guide the generation of gesture sequences that align with both speech content and melodic variations. Furthermore, we decouple emotions from the computation graph and optimize the noise classifier for emotion stylization, reasonably endowing gestures with emotional diversity without damaging the original semantic and melodic information. The main contributions of our work are: (1) We propose ExpGest, to the best of our knowledge, the first motion speaker generation framework under mixed control that combines audio-to-gesture and text-to-motion. (2) We decouple gesture components and introduce a semantic alignment module in the latent space. We separately assign stronger melody relevance and semantic relevance to arms and fingers, generating gesture poses that better express audio content. (3) We introduce a noise emotion classifier into the reverse diffusion process, controlling the emotional style of gestures by optimizing noise gestures through gradient backpropagation. (4) We demonstrate through extensive experiments that the naturalness and richness of generated speaker actions have been improved, surpassing existing methods in terms of motion quality.
\vspace{-20pt}
\section{Approach}
\vspace{-5pt}
Distinct from existing co-speech gesture generation methods~\cite{ao2023gesturediffuclip, yang2023diffusestylegesture} that focus on upper-body or partial movements, \textbf{ExpGest} aims to create expressive talkers with full-body motion corresponding to textual descriptions and co-speech gestures aligned with audio. The overall network framework is shown in Figure~\ref{fig:arc}. We first introduced a unified data representation in Section 2.1. Then, in Section 2.2, we presented a learning framework based on the diffusion model, effectively capturing expressive full-body gestures influenced by both speech and text information through combined text and audio guidance. Finally, in Sections 2.3 and 2.4, we proposed the semantic alignment module and emotion guidance, ensuring that our gesture results conform to semantic descriptions while maintaining emotional expressiveness.\\
\vspace{-20pt}
\subsection{Unified Data Representation}
\vspace{-5pt}
Our method requires the correct preservation of features from different motion datasets, necessitating the unification of various data representations. First, we extract the Euler angles from motion capture data (BVH format) and convert them into rot6D representations corresponding to the 55 joints in SMPL-X~\cite{loper2015smpl}. For 3D locations, we randomly extract a skeleton as a reference and align other data to this skeleton to unify the 3D coordinates. We then scale the root node's displacement based on the extracted skeleton, adjusting the orientation to ensure consistency across all data. Finally, we combine the extracted rot6D rotation representation ($J\times6$), 3D location ($J\times3$), linear velocity ($J\times3$), angular velocity ($J\times6$), and ground contact signals (4) into a kinematic feature representation, where $J$ represents the number of joints. Each frame of the unified motion sequence has a 994-dimensional feature representation, denoted as $x_{0}\in \mathbb{R}^{N\times994}$, where $N$ is the number of motion sequence frames. Due to the lack of mixed-modality data, we artificially synthesized some data. Specifically, we considered the lower body (including the root) below the third spine as the locomotion-involved part, and the remaining upper body as the gesture-involved part. We concatenated the two parts, resulting in 20K artificially generated text-audio-motion matching pairs. We used a ratio of (0.6, 0.4) to mix gesture data and artificial data as the training dataset.\\
\vspace{-20pt}
\subsection{Diffusion Model for Generating Motion Speakers}
\vspace{-5pt}
\begin{figure}
  \begin{center}
  \vspace{-2em}
    \includegraphics[width=1\linewidth]{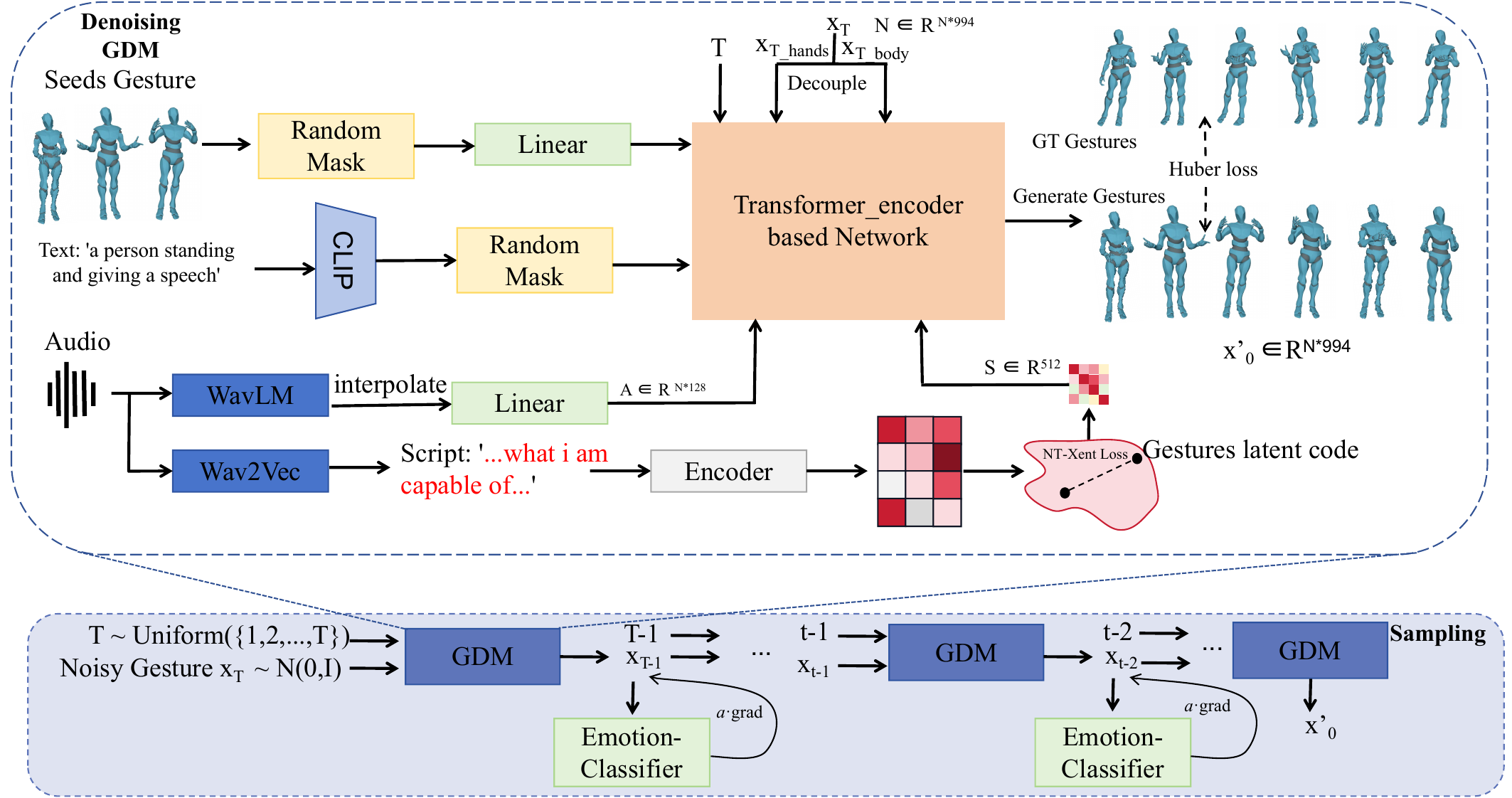}
    \vspace{-1em}
  \end{center}
  \vspace{-15pt}
  \caption{\textbf{Architecture Diagram. }The upper part is the denoising model GDM. Noise step $T$, along with pure Gaussian noise and conditions (text description and audio), is fed into the model as input sequences. The lower part is the sampling step, where we predict $x'_{0}$ through the denoising process and add noise to $x_{t-1}$ via the diffusion process. Subsequently, $x_{t-1}$ is input into the noise emotion classifier to optimize the noise at that step, and the optimized noise is then passed back to the GDM. This cycle continues until t=T becomes t=0.}
  \vspace{-15pt}
  \label{fig:arc}
  \end{figure}
ExpGest builds upon the diffusion model~\cite{ho2020denoising} to generate co-speech gestures. As illustrated in Figure~\ref{fig:arc}, our approach involves learning how to progressively denoise from pure Gaussian noise to obtain expressive full-body gestures. This process encompasses a forward diffusion step, where noise is added to the original gestures, and a subsequent reverse denoising procedure.\\
\textbf{Denoising Diffusion Probabilistic Model.}
We sample clean gestures $x_{0}$ from the distribution of real gestures $x_{0} \sim q(x_{0})$, where $q(x_{0})$ represents the distribution of the original data. Subsequently, Gaussian noise is incrementally added to $x_{0}$. When the number of noise steps $T$ is sufficiently large, the final noisy gesture $x_{T}$ converges to pure Gaussian noise. Each step is modeled as a forward process denoted by $q$. The forward process diffuses the data samples through Gaussian transitions parameterized with a Markov process:
\vspace{-5pt}
\begin{equation}
\begin{aligned}
 q(x_{t}|x_{t-1}) = \mathcal{N}(x_{t};\sqrt{1-\beta_{t}}x_{t-1},\beta_{t}I), \\
 = \mathcal{N}(\sqrt{\frac{\alpha_{t}}{\alpha_{t-1}}}x_{t-1},(1-\frac{\alpha_{t}}{\alpha_{t-1}})I), 
\end{aligned}
\end{equation}
where $\{\beta_{t}\}_{t=1}^{T}$ is the variance schedule and $\alpha_{t} = \prod_{s=1}^{t}(1-\beta_{s})$.\\
Then the reverse process becomes $p_{\theta}(x_{0:T}):=p(x_{T})\prod_{t=1}^{T}p_{\theta}(x_{t-1}|x_{t})$, starting from $x_{T}\sim\mathcal{N}(0,I)$ with noise predictor $\epsilon_{t}^{\theta}$:
\vspace{-5pt}
\begin{equation}
\begin{aligned}
 x_{t-1}= \frac{1}{\sqrt{1-\beta_{t}}}(x_{t}-\frac{\beta_{t}}{\sqrt{1-\alpha_{t}}}\epsilon_{t}^{\theta}(x_{t}))+\sigma_{t}z_{t},\\
\end{aligned}
\end{equation}
where $z_{t} \sim \mathcal{N}(0,I)$ and $\sigma_{t}^{2}=\beta_{t}$ means the variance schedule stays constant.\\
\textbf{Expressive Gesture Diffusion Generator.} 
The objective of ExpGest is to generate motion-capable talking avatars of arbitrary lengths, accompanied by semantically relevant and expressively rich synchronized gestures. However, unlike the original DDPM~\cite{ho2020denoising}, we take into account the extreme physical constraints inherent in 3D human bodies and replace the predicted noise with the original human representation, deviating from image generation. As a result, in each step of our denoising process, we reconstruct the original representation from pure Gaussian noise, ultimately producing the final generated result through a cyclic process of noise addition and denoising:
\vspace{-5pt}
\begin{equation}
\begin{aligned}
    \hat{x}_{0}&= \frac{x_{t}-\sqrt{1-\alpha_{t}} \epsilon_{t}^{\theta}\left(x_{t}|c\right)}{\sqrt{\alpha_{t}}},\\
    x_{t-1}&=\sqrt{\alpha_{t-1}} \hat{x}_{0}+\sqrt{1-\alpha_{t-1}-\sigma_{t}^{2}} \cdot \epsilon_{t}^{\theta}\left(x_{t}|c\right)+\sigma_{t} z_{t},
\end{aligned}
\end{equation}
where $c$ is the conditions. As shown in the upper part of Figure~\ref{fig:arc}, our conditions include noise step $t$, seed posture, motion text, audio information, and semantic latent code. The noise step $t$ and seed posture are projected to the same dimensions through MLP and Linear layers, respectively, and then added together. The audio is encoded using WavLM, and to match the number of gesture frames, it is interpolated in the time dimension. The text description is directly encoded into the CLIP space, while the semantic features are obtained through our semantic alignment module. Finally, we perform Huber loss on $\hat{x}_{0}$ reconstructed at each step and $x_{0}$ sampled in the real distribution to optimize ExpGest:
\vspace{-5pt}
\begin{equation}
\begin{aligned}
 \mathcal{L}_{Huber}=E_{x_{0}\sim q(x_{0}|c),t\sim[1,T]}[HuberLoss(x_{0}-\hat{x}_{0})],\\
\end{aligned}
\end{equation}
It is worth noting that during the denoising process, if there is only a single modality input, the other modality is directly masked and added to the conditions in the form of $\varnothing$.
\vspace{-10pt}
\subsection{Semantic Alignment in Latent Space}
\vspace{-5pt}
\begin{figure}
  \begin{center}
  \vspace{-2em}
    \includegraphics[width=1\linewidth]{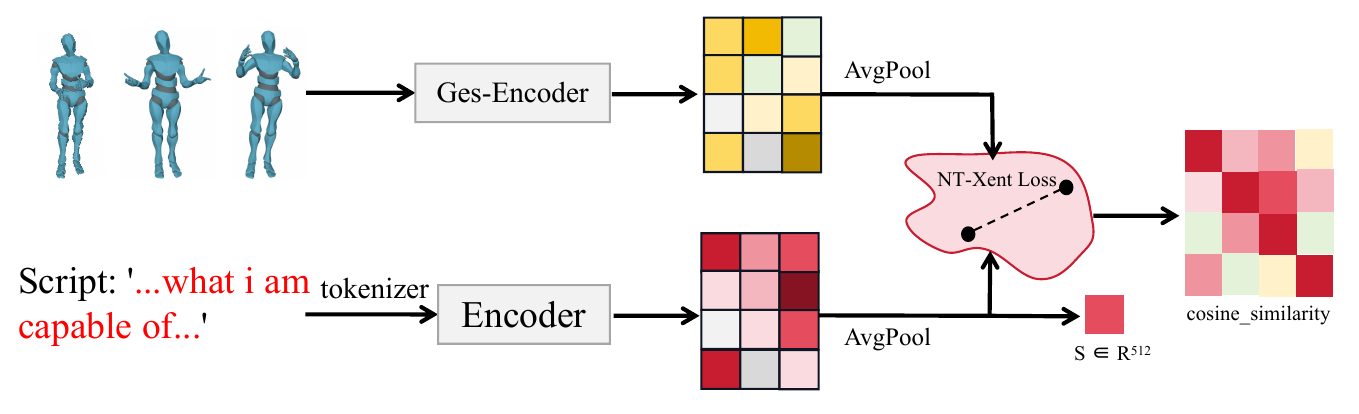}
    \vspace{-1em}
  \end{center}
  \vspace{-20pt}
  \caption{\textbf{Semantic Alignment Module.} We employ contrastive learning to encode gestures and audio semantics into a shared latent space and achieve alignment in the latent space.}
  \vspace{-15pt}
  \label{fig:sem}
  \end{figure}
In previous gesture generation methods, the crucial aspect of semantic alignment was often overlooked, posing a significant challenge for accurately generating semantically aligned actions due to the many-to-many mapping relationship between audio content and gesture sequences. To tackle this issue, we learn a joint embedding space for gestures and audio transcriptions, allowing their alignment in latent space and uncovering semantic associations between both modalities. As depicted in Figure~\ref{fig:sem}, we initially train gesture and transcription encoders using a motion-VAE structure and a BERT tokenizer~\cite{devlin2018bert}, respectively. We parameterize transcriptions into tokenized word embedding sequences and linearly map them to a space with the same dimensions as the gesture latent code. We employ global average pooling to extract global information from both modalities and utilize CLIP-style contrastive learning to fine-tune the encoders. The NT-Xent~\cite{chen2020simple} Loss serves as the objective to maximize the similarity of transcription-gesture matching pairs in latent space while minimizing the similarity of non-matching pairs. Formally, the loss function is as follows:
\begin{equation}
\begin{aligned}
    \mathcal{L}(s,g) = -log\frac{exp(sim(z_{s},z_{g})/\tau)}{\sum_{k\in K}exp(sim(z_{s},z_{k})/\tau)},\\
\end{aligned}
\end{equation}
where, $z_{s}$ and $z_{g}$ are the latent space representations of a matching transcription-gesture pair. $sim$ is similarity score between two latent codes, $K$ is a set containing one positive sample transcription and a group of negative sample gestures, and $\tau$ is the temperature parameter used to adjust the sensitivity of the function. Finally, we freeze the trained semantic alignment module and deploy only the transcription encoder into the GDM, ensuring that the final generated results accurately capture the semantic content.
\vspace{-10pt}
\subsection{Noise-based Emotion Guided Classifier}
\vspace{-5pt}
Previous methods transformed emotions into a set of one-hot encodings embedded directly into the conditions, controlling emotions by appending identity. However, this approach fails to capture the underlying relationships and continuity between emotions, making it difficult for the model to learn smooth transitions between different emotions. To address this issue, we introduce a noise emotion classifier that optimizes the denoised noise gesture at each sampling step, moving it towards a human-specified emotional direction. This method is decoupled from the diffusion computation graph, preserving the original semantic content and making it applicable to different structures in the same representation. Specifically, we randomly select emotion-gesture matching pairs and add Gaussian noise of random steps to the gestures, turning them into noisy gestures. We then train the noise emotion classifier based on the noisy gesture-emotion matching pair data. After training, we graft the classifier onto the GDM, input the denoised $x_{t}$ into the GDM, compute the gradient with the specified emotion, and then backpropagate it to $x_{t}$ for optimization. The formula is as follows:

\begin{equation}
\begin{aligned}
    \hat{x}_{t}=x_{t}+\alpha \cdot \nabla_{x_t} \mathcal{L}(\text{Emo\_Classifier}(x_{t}),y)\\
\end{aligned}
\end{equation}

where $\alpha$ is gradient update weight and $y$ is the real emotion label.
\vspace{-10pt}
\section{Experiments}
\vspace{-5pt}
In this section, we discuss ExpGest's technical details, evaluation, comparison with other frameworks, potential applications, and design choices validation through ablation studies.\\
\textbf{Implementation Details.} We first downsample all data to 20 FPS to unify the frame rate. For text-motion data, we randomly truncate long-frame data to a motion length of 180 frames, and for short-frame data, we select a length of 60 to 180 frames, padding with zeros to extend to a uniform 180-frame length. We set the maximum text length for CLIP encoding to 20. For audio-gesture data, we randomly truncate the audio and corresponding gesture sequence to 180 frames. For artificially synthesized data, we only use motions with a large displacement for the lower body (including running, walking, jumping, standing, sitting, etc.) and replace the gesture data below the third spine with motion data, removing some unnatural sequences. For the semantic alignment module, we align transcription texts and gesture motions in a shared embedding space using a VAE encoder and a BERT tokenizer. For the noise emotion classifier module, we project the noise gesture motion representation and predict 8-bit one-hot encoded emotion categories. We construct a diffusion model using a 12-layer Transformer encoder and set the diffusion steps to 1000. We train the model for approximately 72 hours on 4 NVIDIA Tesla V100 GPUs, totaling 800K steps.
\vspace{-10pt}
\subsection{Dataset and Evaluation Metrics}
\vspace{-5pt}
\textbf{Datasets.} Our audio data comes from the large-scale, high-quality \textbf{BEAT} dataset~\cite{liu2022beat}, featuring 76 hours of multimodal speech data from 30 speakers in 8 emotions and 4 languages. We selected the \textbf{English} data from both presentations and conversations for training. We also used \textbf{AMASS}\cite{mahmood2019amass} and \textbf{100-STYLE}\cite{mason2018style} for locomotion training. From AMASS, we selected motion pairs with text descriptions, containing only the body's SMPL representation and zero-filled hand representations. 100-STYLE, a multi-style motion dataset, includes various actions each with a BVH file and short phrase description. We processed these data into the motion representation described in Sec 2.1 for unified training.\\
\textbf{Evaluation Metrics.} 
We use three metrics to evaluate the quality, emotion alignment, and semantic matching of gestures generated by ExpGest. Gesture quality is assessed using Fréchet Gesture Distance (FGD)~\cite{yoon2020speech}, which calculates the distance between latent feature distributions of generated and real gestures, evaluating gesture quality. Lower FGD values indicate higher motion quality. Additionally, we propose a Semantic Alignment (SA) metric to assess the semantic consistency between generated gestures and audio. We use the trained VAE encoder to encode transcribed audio text and gestures into a shared latent space and evaluate their pre-established similarity as the assessment score.The calculation formula is as follows:
\begin{equation}
\begin{aligned}
    SA = cos(avg\_pool(V_{g}(G_{pred})),avg\_pool(V_{s}(S))),\\
\end{aligned}
\end{equation}
where \textit{G} represents the gestures generated by the model, and \textit{S} denotes the hidden states encoded by the BERT~\cite{devlin2018bert} model after tokenizing the transcribed text, serving as a representation of the semantics. We also introduce Emotion Alignment (EA) and Emotion Control Success Rate (EC) to evaluate the model's stylization capabilities. Unlike in Section 2.4, we pre-train a gesture emotion classifier, which doesn't require adding noise to the gesture motion but directly classifies emotions in the original gestures. EA represents consistency with the true emotions of the audio, while EC denotes the success rate of style transfer for specified emotions. Higher values for both EA and EC indicate more accurate emotion guidance.
\begin{equation}
\begin{aligned}
    E_{C/A} = \frac{Emo\_Classifier(G_{pred},Emotions)}{Sum(Emotions)},\\
\end{aligned}
\end{equation}

\begin{table}
\centering
\begin{adjustbox}{scale=0.77}
\begin{tabular}{lccccccccc}
\toprule[1.5pt]
\multirow{1}{*}{Method} & \multicolumn{1}{c}{$FGD (Raw)$} &\multicolumn{1}{c}{$FGD (Feature)$} &\multicolumn{1}{c}{$EA$}  &\multicolumn{1}{c}{$SA$}\\ \hline
Ground Truth & 0.0 &0.0 &0.97&0.91 \\ 
CAMN~\cite{liu2022beat} &52.4 &263.9 &- &-\\ 
Trimodal~\cite{yoon2020speech} &47.6 &212.7 &- &-\\ 
DiffStyleGesture~\cite{yang2023diffusestylegesture} &33.7 &133.9 &0.60 &0.11 \\ \hline
ExpGest &\textbf{11.7} &\textbf{76.6} &\textbf{0.91} &\textbf{0.61}\\
ExpGest w/ hybrid-gudied &25.4 &129.3 &- &-\\
\hline
\end{tabular}
\end{adjustbox}
\caption{\textbf{Quantitative Assessment Outcomes.} Boldface denotes the top-performing method for each evaluation criterion. All experiments are conducted and appraised on the BEAT dataset. Trimodal~\cite{yoon2020speech} and CAMN~\cite{liu2022beat} outcomes are reproduced utilizing the official code, while DiffStyleGesture~\cite{yang2023diffusestylegesture} results are retrained on the BEAT~\cite{liu2022beat} dataset employing the official training approach. A dash (``-") signifies the unavailability of a particular metric.}
\label{tab:qua}
\end{table}

\subsection{Comparison to State-of-the-art Methods} 
\begin{figure}[b]
  \begin{center}
  \vspace{5pt}
    \includegraphics[width=1\linewidth]{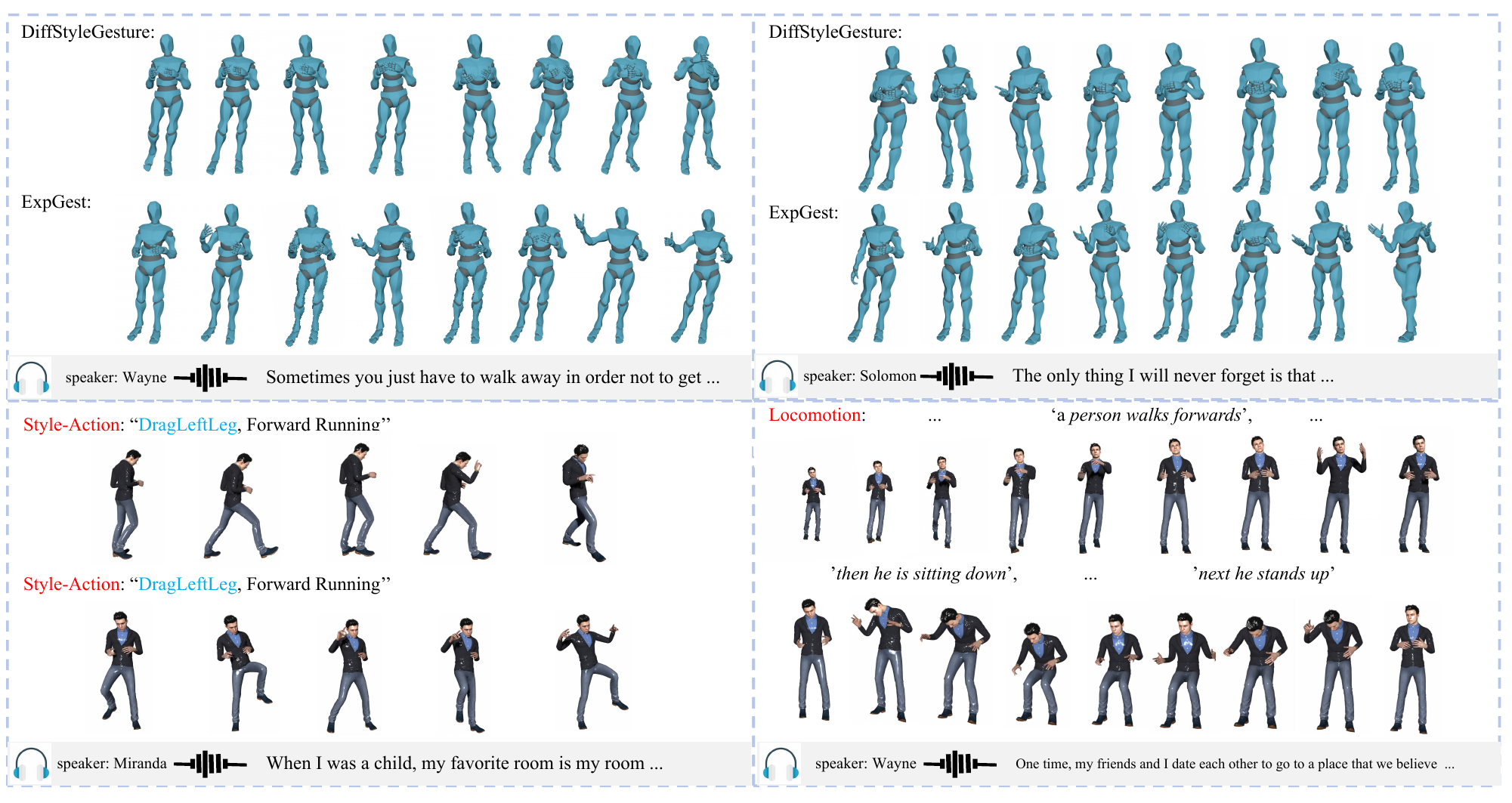}
  \end{center}
  \caption{The first row compares our method's generative performance in an audio-guided scenario with state-of-the-art techniques, yielding more expressive gestures. The second row demonstrates a speaker's generation with rich movements, guided by audio combined with action or text. See \textbf{Supp} for more videos.}
  \label{fig:qual}
  \end{figure}
  
We compare our method with three cutting-edge approaches: Trimodal~\cite{yoon2020speech}, CaMN~\cite{liu2022beat}, and DiffStyleGesture~\cite{yang2023diffusestylegesture}. We remove the text guidance describing motion and use an audio-only mode for comparison. Quantitative results in Table~\ref{tab:qua} show our method consistently outperforms all other methods across all evaluations. Specifically, in terms of Fréchet Gesture Distance, we achieve state-of-the-art performance in the audio-only mode. Our method improves the feature space by 57.3 (42.7\%) and the raw space by 22 (65.2\%) in evaluations with only upper-body limbs compared to our baseline. Although there is a slight decrease in the mixed mode, our method still outperforms the baseline. Thanks to the semantic-gesture joint embedding space and using aligned semantic features as guidance conditions, our model achieves improvements in SA. Furthermore, by optimizing the noise gradient and guiding generation towards specified emotion styles, our method significantly outperforms existing methods in EA and EC metrics.
\begin{table}
\centering
\begin{adjustbox}{scale=0.9}
\begin{tabular}{lccccccccc}
\toprule[1.5pt]
\multirow{1}{*}{Method} & \multicolumn{1}{c}{$EA$} & \multicolumn{1}{c}{$EC$} & \multicolumn{1}{c}{$EA_{hands}$} &\multicolumn{1}{c}{$EC_{hands}$}\\ \hline
Ground Truth & 0.97 &- &0.95 &-\\ 
DiffStyleGesture~\cite{yang2023diffusestylegesture} & 0.60 & 0.27 &0.49 &0.19 \\ \hline
DiffStyleGesture w/ EG & 0.83 &0.69 &0.70 &0.63 \\
ExpGest  & \textbf{0.91} &\textbf{0.83} &\textbf{0.81} &\textbf{0.70} \\
\hline
\end{tabular}
\end{adjustbox}
\caption{\textbf{Emotion-guided Results.} The bold values represent the best results. All methods are trained on the same dataset.}
\label{tab:emo}
\end{table}

\begin{table*}
\centering
\begin{adjustbox}{scale=1.0}
\begin{tabular}{lccccccccc}
\toprule[1.5pt]
\multirow{1}{*}{Method} & \multicolumn{1}{c}{\textit{Human-likeness}} & \multicolumn{1}{c}{\textit{Gesture-appropriateness}} & \multicolumn{1}{c}{\textit{Emotion-compatibility}} &\multicolumn{1}{c}{\textit{Global-coherence}}\\ \hline
Ground Truth & $4.61 \pm 0.17$ & $4.72 \pm 0.20$ &$3.62 \pm 0.27$ &-\\ 
DiffStyleGesture~\cite{yang2023diffusestylegesture} & $3.46 \pm 0.16$ & $3.61 \pm 0.11$ & $1.22 \pm 0.19$ & - \\ \hline
ExpGest (audio-only) & $4.17 \pm 0.13$ &$4.20 \pm 0.18$ &$3.88 \pm 0.21$ & - \\
ExpGest (audio-action) & $3.76 \pm 0.08$ & $3.83 \pm 0.11$ & $2.63 \pm 0.09$ &$3.61 \pm 0.17$\\
ExpGest (audio-text) &$3.72 \pm 0.10$ & $3.91 \pm 0.15$ &$2.27 \pm 0.06$ &$3.47 \pm 0.16$    \\ 
\hline
\end{tabular}
\end{adjustbox}
\caption{\textbf{95\% Confidence Interval for User Study Average Score.} Since the GT data from the BEAT dataset lacks global locomotion, we only compare gesture sequences generated by ExpGest in audio-only mode with GT and DiffStyleGesture, using user ratings. For the Global-coherence metric, we combine action or text descriptions with audio as input and have users evaluate the overall coordination of the generated speaker motion.}
\label{tab:user}
\end{table*}
\subsection{Is the emotion-guided classifier helpful?}
To evaluate our emotion classifier's guidance capability, we randomly select two to four emotions from the eight available in the BEAT~\cite{liu2022beat} dataset and assign them to each audio segment in the validation set. We feed the generated results into a pre-trained emotion classifier to determine the emotion category and compare the output with the ground truth emotion labels. As shown in Table~\ref{tab:emo}, our baseline method significantly outperforms DiffStyleGesture~\cite{yang2023diffusestylegesture} in emotion-alignment and emotion-controllability metrics. We also provide an emotion evaluation metric for fingers, observing noticeable changes in some emotions. For instance, under the angry emotion, the virtual character often displays pointing gestures while speaking. To substantiate our emotion classifier's effectiveness, we incorporate it into the DiffStyleGesture method's inference stage for evaluation. Using the same data representation ensures a fair comparison with consistent evaluation logic and classifier model across both methods. Table~\ref{tab:emo} shows that, with the emotion classifier, DiffStyleGesture achieves higher emotion alignment compared to the original approach (relying on one-hot encoding for emotion representation).
\subsection{User Study and Results}
\textbf{User Study.} To evaluate ExpGest's real-world visual performance, we conducted a user study comparing generated gesture sequences under various guidance modes, DiffStyleGesture, and ground truth data. We curated 40 audio clips from the BEAT test set, featuring diverse voices, and divided them into short (8-16 seconds) and long (60-90 seconds) segments. About 120 participants evaluated the gesture sequences rendered into dynamic virtual speakers using Blender. We employed four evaluation metrics: ``Human-likeness", ``Gesture-appropriateness", ``Emotion-compatibility", and ``Global-coherence", each scored on a 1-5 scale. Table~\ref{tab:user} presents the average scores, showing ExpGest outperforms previous methods in visual performance and gains higher acceptance from participants. The mixed guidance approach demonstrates remarkable synergy in generated gestures.\\
\textbf{Results.} Figure~\ref{fig:qual} (top) demonstrates a comparison of ExpGest with prior state-of-the-art methods under single audio guidance. We uniformly selected serene recordings from Wayne and Solomon, two speakers in the BEAT dataset, as test audio. The generation results of DiffStyleGesture, which relies solely on melody features, are relatively static with little variation. In contrast, ExpGest, benefiting from semantic alignment in the latent space, reveals highly expressive gesture results. Furthermore, our proposed noise emotion guide can naturally and smoothly perform emotional transitions, thereby enhancing the expressiveness of the generated results (see \textbf{Spp video} for reference). Figure~\ref{fig:qual} (bottom) showcases the superior performance of ExpGest under mixed control. By integrating audio with action, ExpGest can generate a variety of motion sequences, such as ``dragging left leg, running forward". Simultaneously, using existing large language models to edit the action text description for a speech (``he looks for a place to sit while talking"), combined with the audio itself, a vivid speaking character can be generated. 

\subsection{Ablation Studies}
\begin{table}
\centering
\begin{adjustbox}{scale=0.8}
\begin{tabular}{lccccccccc}
\toprule[1.5pt]
\multirow{1}{*}{Method} & \multicolumn{1}{c}{\textit{$FGD$ (raw)}} & \multicolumn{1}{c}{\textit{$FGD$ (feature)}} & \multicolumn{1}{c}{\textit{SA}} & \multicolumn{1}{c}{\textit{EA}}\\ \hline
DiffStyleGesture~\cite{yang2023diffusestylegesture} & 33.7 & 133.9 & 0.11 &0.60 \\ \hline
Baseline  & 25.9 & 95.5 & 0.14 &0.57  \\
Baseline w/ EG  & 28.2 & 115.3 & 0.10 &0.89  \\
Baseline w/ SA  & 18.6 & 89.1 & 0.63 & 0.62\\
ExpGes  & 11.7 & 76.6 & 0.61 &0.91    \\ 
\hline
\end{tabular}
\end{adjustbox}
\caption{\textbf{Ablation study.} Demonstrating the effectiveness of semantic alignment module and emotion guidance module.}
\label{tab:abl}
\end{table}
we designed the following ablation experiments. The results are detailed in Table~\ref{tab:abl}. Our baseline method has already achieved significant improvements compared to DiffStyleGesture. Subsequently, we incorporated an Emotion-Guided Classifier (EG) during the reverse diffusion process, optimizing noise using gradient backpropagation. Although the performance in FGD has decreased, it still outperforms previous methods and significantly improves emotion alignment. Then, we introduced the Semantic Alignment (SA) module. Compared to the baseline, FGD has increased significantly, indicating the importance of semantic alignment in gesture generation, and since there are no additional optimization operations in reverse diffusion, the generation speed is also close to DiffStyleGesture, taking approximately 20 seconds to generate a 180-frame gesture slice. In this experiment, we conducted ablation studies only on the proposed baseline, without involving mixed guidance. All evaluation results were generated using pure audio guidance.
\vspace{-10pt}
\section{Conclusions}
\vspace{-5pt}
In this paper, we propose ExpGest, a novel framework using the diffusion generation model to create motion speakers from mixed audio-text guidance. Training on heterogeneous gesture-audio and text-motion pairs allows effective capture of human motion nuances. Our approach generates natural-looking speaker gesture sequences, laying the groundwork for large-scale motion speaker generation with potential applications in virtual agents, movies, and human-computer interaction. In the future, we will incorporate existing action capture methods~\cite{cheng2023bopr, liang2024ropetp} to generate more diverse continuous natural data~\cite{yu2024signavatars}, so as to achieve more realistic training effects.

\bibliographystyle{IEEEbib}
\small
\bibliography{icme2023template}
\end{document}


\sloppy

\def\x{{\mathbf x}}
\def\L{{\cal L}}

\title{ReBaR: Reference-Based Reasoning for Robust Human Pose and Shape Estimation from Monocular Images \\ Appendix}
%
\name{Anonymous ICME submission}
\address{}
\maketitle

\section{Appendix}
\label{sec:Appendix}
This supplement provides additional experimental results to enhance the main paper. In Section ~\ref{sec:sec1}, we offer more implementation details. In Section ~\ref{sec:sec2}, we present additional experimental results. In Section ~\ref{sec:sec3}, we use more in-the-wild data to compare and verify the effectiveness of ReBaR on extremely challenging real-world problems.
\section{More Details}

In our experiments, our backbone network is initialized by the HRNet model weights pre-trained on the MPII~\cite{andriluka20142d} dataset for 2D key point detection tasks.

\begin{figure}
  \centering
  \includegraphics[width=1\linewidth]{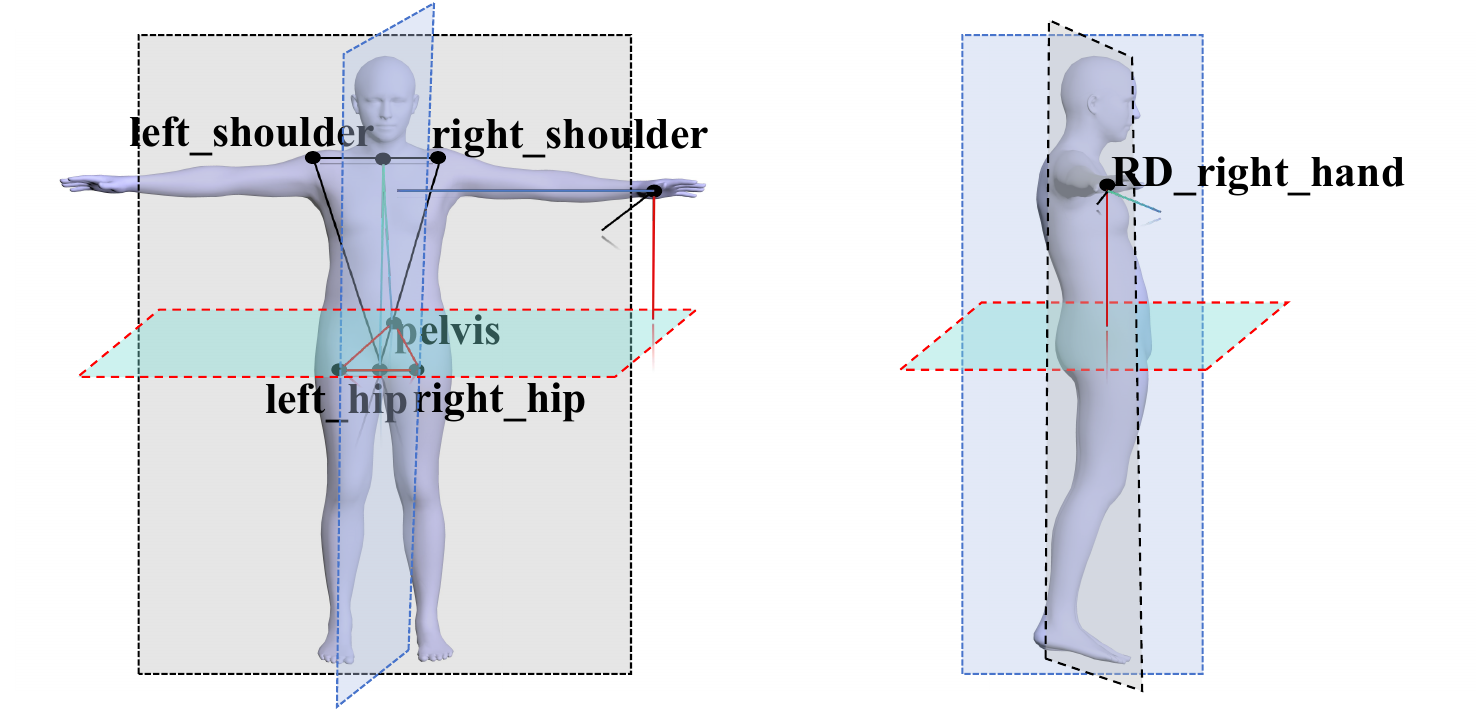}
  \caption{\noindent\textbf{Illustration of the calculation of the relative depth from the torso plane.} Shows the torso points for constructing the torso planes and the relative depth of the right hand.}
  \label{fig:rp}
\end{figure}

\noindent \textbf{Torso Plane and Relative Depth}\\
As shown in figure~\ref{fig:rp}, we select the shoulder, hip, and pelvis points on the human torso to establish the frontal, lateral, and transverse planes of the torso. The center points of the skin regions of each joint are defined as the actual joint positions, and joint axes are defined by bringing equations greater than 0 into the plane as positive directions, and the depth value is calculated based on the distance from the point to the plane. This is used as a basic fact label for relative depth, and explicit plane consistency constraints are constructed.
\noindent \textbf{Loss Weight} \\ The weight distribution of our supervision functions is as follows: segmentation map supervision, reference feature 2D/3D auxiliary constraints, and relative depth supervision all have a weight of 60, while the SMPL parameter and keypoint supervision are five times greater than the former. This balanced approach contributes to the model's effectiveness.

\noindent \textbf{Data Augmentation} \\In all the experiments in the paper, except that Table 2 does not use any augmentation on the AGORA dataset, other experiments use the same data augmentation method as PARE.

\noindent \textbf{Auxiliary Loss}\\ In the process of training ReBaR, we respectively use global 2D/3D keypoints, body/part attention map and relative depth supervision to assist the network to learn body-aware part features. For the supervision of body/part attention, we only supervise on the COCO-EFT dataset, and reset the weight of $L_{x\_seg}$ in the loss function to 0 on the mix and 3DPW datasets.For the AGORA dataset, we do not supervise the relative depth.
\label{sec:sec1}

\noindent \textbf{Inference Time and parameter Counts}\\
Our model employs a Transformer to trade computational cost for exceptional performance, resulting in a longer inference time compared to PARE and CLIFF, but still within an acceptable range. As shown in Table~\ref{tab:iftime}, we tested the inference speed for a single image using the same GPU, T4-8C, and calculated the total number of parameters for the PARE, CLIFF, and ReBaR models. Additionally, we recorded the model training costs in the log files, with training conducted on four V100 GPUs. PARE takes 16 hours to train with the original settings (72 hours on a 2080Ti), while ReBaR requires 8 hours for training on COCO and 20 hours for fine-tuning on the mixed dataset. Lastly, when evaluating AGORA, we fine-tuned the model for 10 hours on a single V100 GPU using the training set. CLIFF does not provide open-source training code, so it is not included in the statistics.

\noindent \textbf{Part/Body segmentation maps}\\ To obtain segmentation maps for the auxiliary supervision of the body attention map and part attention map in the AGE module, segmentation labels are required for each part of the original image. However, labeling the original image is a time-consuming and costly process. Luckily, we can utilize the SMPL model to obtain the vertex coordinates in the camera coordinate system that correspond to each image. Using weak perspective transformation techniques~\cite{kissos2020beyond}, we can then generate the 2D vertex coordinates in the pixel coordinate system that are needed to generate the segmentation map. This method enables us to obtain the necessary segmentation maps without the need for costly and manual labeling of the original image. We introduce inverse weak-perspective projection to generate body/part attention map labels. First, let's discuss weak-perspective projection:\\
\begin{table}
\centering
\begin{adjustbox}{scale=0.7}
\begin{tabular}{lcccccccc}
\toprule[1.5pt]
{Method} & {PARAMS} & {Inference} & {Training} & {FLOPS} & {$MJE$} & {$MJE_{z}$}\\ 
\hline
PARE & 125.5MB & 54ms  &16h  & 14.9G &74.5 &58.2\\ 
PARE + TF & 155.8MB & 62ms  &19h  & 18.1G &73.2 &56.6\\
CLIFF* & 305MB & 111ms  &-  & - &69.0 & 51.8\\
\hline
ReBaR (Ours) & 222MB & 83ms  &28h  & 26.4G &69.1 &46.6\\
ReBaR* (Ours) & 361.1MB & 132ms  &45h  & - &67.2 &45.5\\
\bottomrule[1.5pt]
\end{tabular}
\end{adjustbox}
\caption{\textbf{More details about the model.} * means using HRNet-48. TF means using Transformer Encoder.}
\label{tab:iftime}
\end{table}

\begin{figure}
  \centering
  \includegraphics[width=1\linewidth]{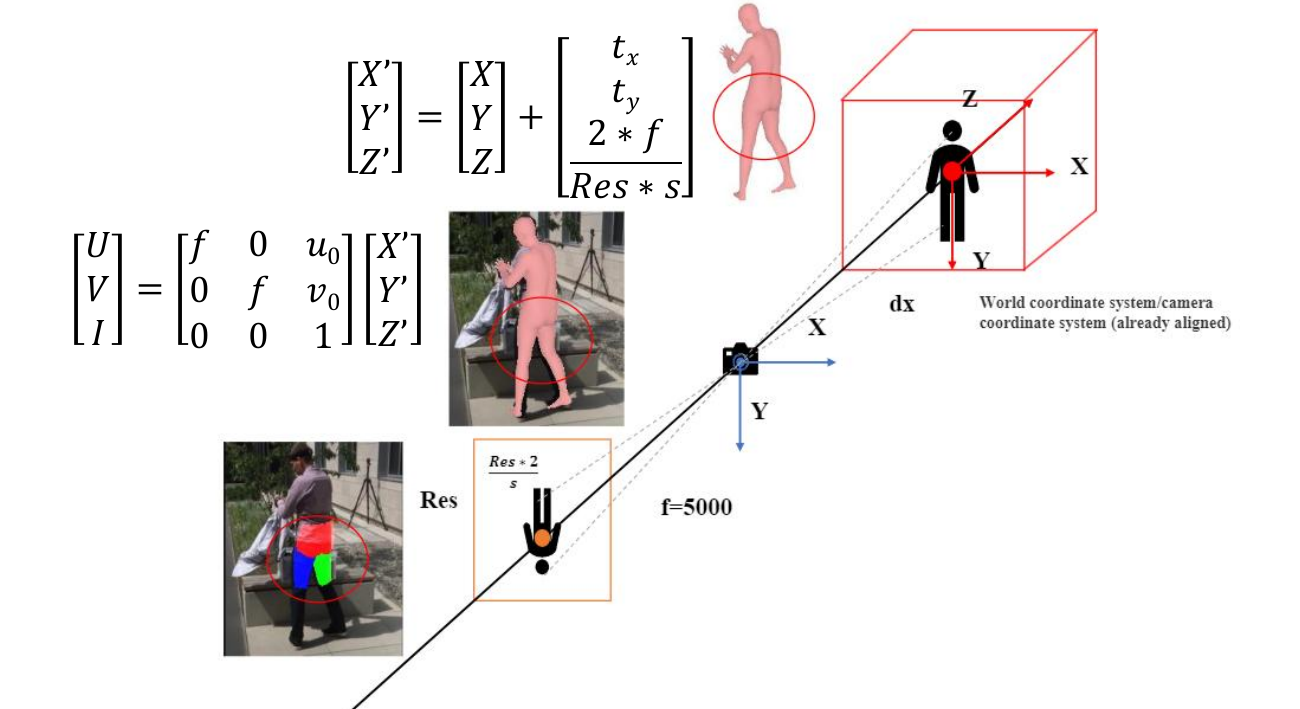}
  \caption{\noindent\textbf{Weak-perspective projection illustration.}}
  \label{fig:weakpp}
\end{figure}

The weak-perspective projection, as described in ~\cite{kissos2020beyond}, is based on the premise that the focal length and object distance are sufficiently large, allowing for the neglect of variations in the object along the Z-axis. It is a technique for converting three-dimensional camera coordinates into pixel coordinates. Before performing the projection, the 3D keypoints are normalized to a cube in the range [-1, 1], and the camera is aligned to the world coordinate system origin (the cube's center). Next, the projection camera parameters ($s, t_{x}, t_{y}$) are required, where s represents the human body's scale in the cropped image, and $t_{x}$ and $t_{y}$ represent the translation within the cube. In this study, the cropping size Res is 224. To perform the weak-perspective projection, we first increase the object distance, as shown in the following formula:
\begin{equation}
\begin{aligned}
  t_{z}=\frac{2 \times f}{Res \times s}
\end{aligned}
\end{equation}

where, f represents the focal length, which is set to 5000 in our study. Subsequently, we add $t_{x}$, $t_{y}$, and $t_{z}$ as translations to the 3D keypoints and introduce the following formula for weak-perspective projection:

\begin{equation}
\begin{aligned}
\quad
\begin{bmatrix} U\\V\\I \end{bmatrix}=\begin{bmatrix} f & 0 &u_{0} \\0&f&v_{0}\\0&0&1 \end{bmatrix} \begin{bmatrix} X+t_{x}\\Y+t_{y}\\Z+t_{z} \end{bmatrix}
\quad
\end{aligned}
\end{equation}

where we set $u_{0}$ and $v_{0}$ to the image's center point position, and f represents the focal length. The resulting $U$ and $V$ are the 2D keypoints in the pixel coordinate system after projection. In the reverse process, we obtain 3D mesh points and 3D/2D keypoints using the ground truth (GT) labels and the SMPL model. We then compute the translation in reverse using the 3D/2D keypoints. Subsequently, we add the translation to the 3D mesh points in the same manner and perform weak-perspective projection to obtain body/part attention map labels. The process is shown in Figure~\ref{fig:weakpp}.
\section{Experiments}
\begin{figure}
\centering
  \includegraphics[width=1\linewidth]{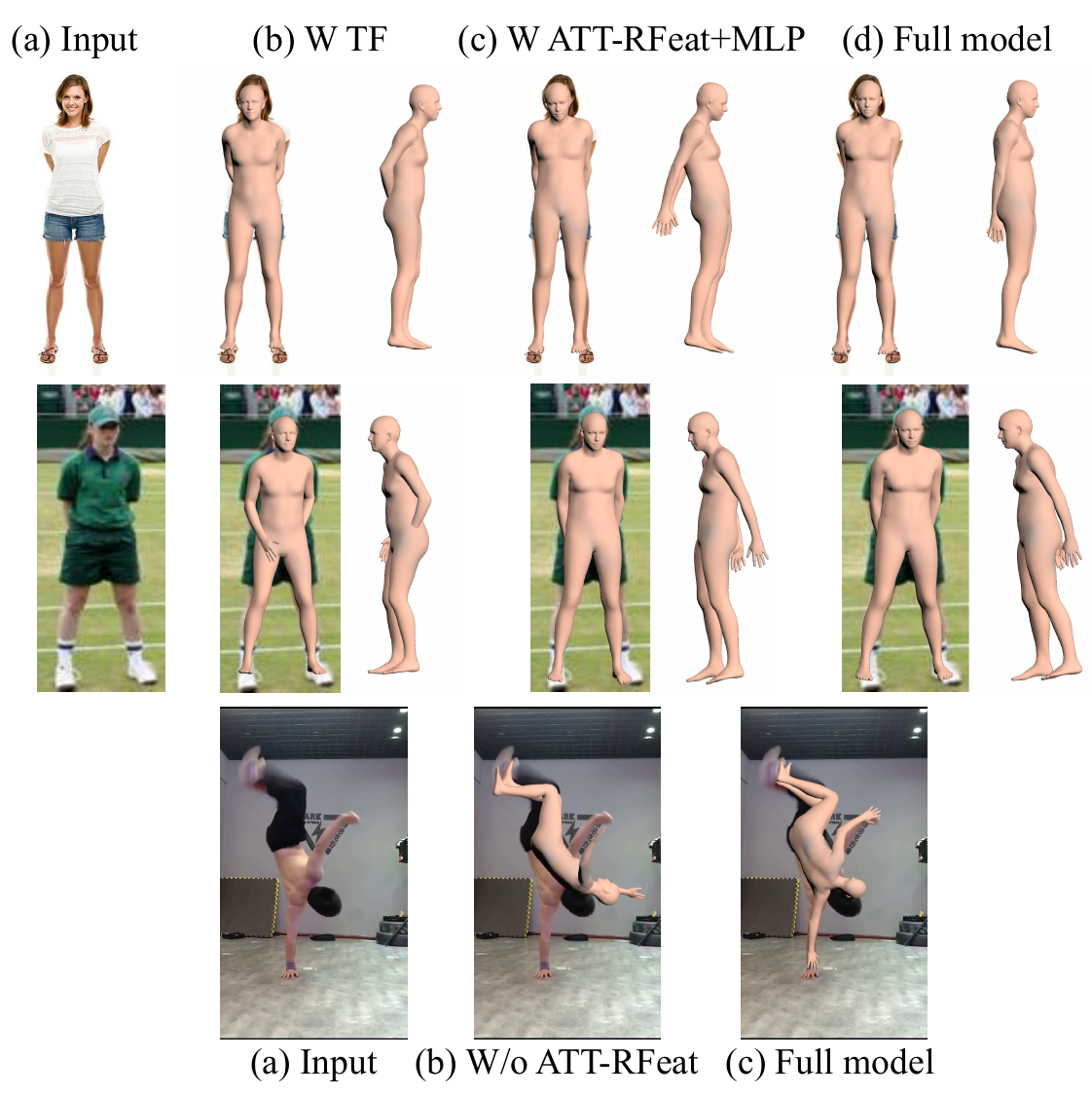}
  \caption{\noindent\textbf{The role of BAR module and ATT-RFeat.} For the first two rows, from left to right: input image, PARE+transformer, PARE+BAR, and the result of the full model. For the last line, from left to right: input image, w/o ATT-Rfeat, and the result of the full model.}
  \label{fig:bar}
\end{figure}

\begin{figure*}
  \centering
  \includegraphics[width=1\linewidth]{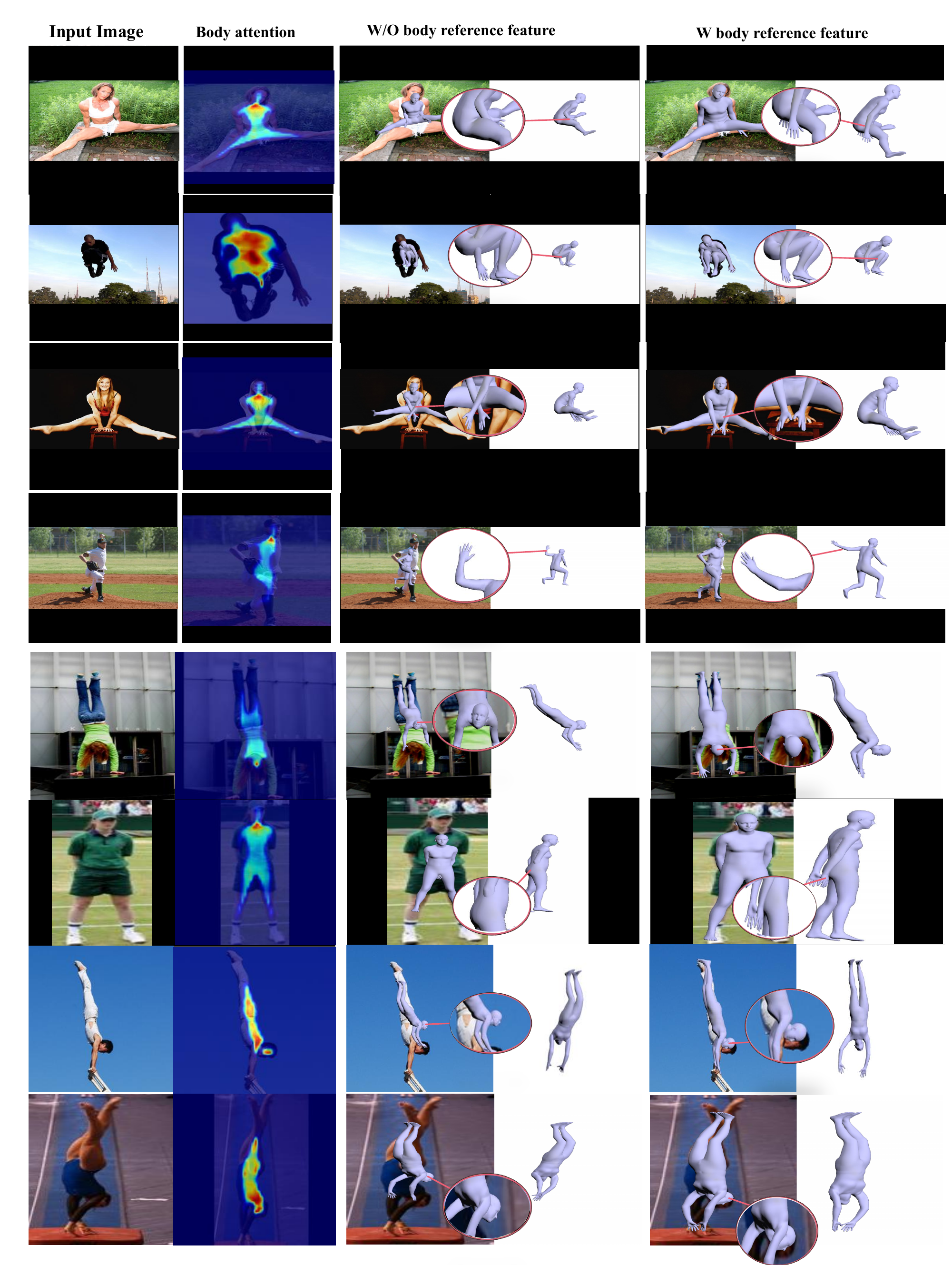}
  \caption{\noindent\textbf{The role of Attention-guided reference feature.} From left to right: input image, body attention map, the result of discarding Attention-guided reference feature, and the result of the full model.}
  \label{fig:dep}
\end{figure*}

In this section, we supplement more experimental results and ablation experiments to verify the effectiveness of ReBaR.
\label{sec:sec2}
\noindent \textbf{Analysis of BAR Module}\\
BAR is the core part of ReBaR. Its role is to construct body perception component features and establish connections between components, so as to infer the posture of occluded parts through more reasonable visible information. This also makes ReBaR have a better ability to alleviate the depth blur problem than existing methods. As shown in Figure ~\ref{fig:bar}, we directly associate the component features of PARE and cannot correctly infer the correct posture of the occluded forearm (such as the wrong posture of the forearm bending forward), but after adding the body reference condition, the model infers Relatively correct posture, although its depth information is not very accurate, and then with the help of relative torso depth constraints, ReBaR accurately infers the posture of the arms behind it.\\
In addition, we also found that in some extremely challenging actions, the body reference condition can greatly improve the stability of the root node. As shown in Figure ~\ref{fig:bar}, in the inverted action, because PARE independently predicts the root joint, it causes an error in the global rotation direction. However, our method ReBaR accurately predicts a reasonable direction through the root feature of body perception, avoiding the problem of random rotation of the human body in video capture.\\
\noindent \textbf{The Role of Attention-guided Reference Feature}\\
Figure~\ref{fig:dep} illustrates the qualitative improvement of our method in challenging cases such as severe occlusion, challenging poses, and depth ambiguity. This demonstrates the importance of body-aware part features encoding and utilizing visual information around parts and Attention-guided reference feature to address depth ambiguity and self-occlusion issues.
\begin{table}
\centering
\begin{tabular}{lcccc}
\toprule[1.5pt]
\multirow{2}{*}{Method} & \multicolumn{3}{c}{3DPW-All}  \\ 
 & MJE$\downarrow$ & PAMJE$\downarrow$ & V2V$\downarrow$  \\ \hline
Baseline~\cite{kocabas2021pare} & 91.0 & 56.7  &108.2\\ 
Baseline w TF & 91.5 & 55.7  &108.6\\ 
Baseline w HMR-RFeat+TF & 90.1 & 55.0 &106.1 \\\hline
ReBaR w/o $L_{3D}$ &89.4 &55.8 &106.7\\
ReBaR w/o $L_{2D}$ &90.6 &54.8 &107.2\\
ReBaR w/o $L_{RD}$ &89.0 &54.2 &105.7\\
ReBaR &\textbf{88.6} &\textbf{53.7} &\textbf{105.3}\\ \bottomrule[1.5pt]
\end{tabular}
\caption{\textbf{Ablation study of ReBaR on 3DPW.} All methods are trained on COCO-EFT-PART.}
\label{tab:aba}
\end{table}
\noindent \textbf{Ablation Experiments With Attention-guided Reference Features} \\To evaluate the effectiveness of the body-aware regressor module, we conducted a set of comparative experiments. We used the global feature of the HMR model as a reference and directly associated it with the part feature to regress the SMPL model parameters using the same transformers. The results, as shown in Table~\ref{tab:aba}, indicate that the global reference encoding only marginally improves the PAMJE metric compared to PARE. However, when we replaced the HMR feature with the Attention-guided reference feature of the AGE module, the model's performance significantly improved. This demonstrates the importance of using Attention-guided referenced feature for inferring part poses.\\
\noindent \textbf{Ablation Experiments With Auxiliary Constraints} \\In this experiment, we use HR-W32 as the backbone and train all comparison methods on the small COCO-2014-EFT (22K) dataset (a subset of the COCO-EFT dataset). We first combine PARE with Transformers and evaluate the performance on the 3DPW dataset. Table~\ref{tab:aba} shows that integrating both techniques slightly reduces PAMJE but increases MJE and PVE. In contrast, our proposed ReBaR significantly improves PARE, indicating that learning Attention-guided reference feature substantially contributes to the performance gain. Furthermore, we validate the auxiliary constraints, i.e., 2D keypoints $L_{2D}$, 3D keypoints 
 $L_{3D}$, and relative depth $L_{RD}$, which bring different levels of improvement to our proposed Attention-guided reference feature. Compared to the unconstrained HMR-feature and single-information supervised body-feature, establishing dependencies between 2D and 3D information in space can better construct stable reference conditions, thereby greatly improving joint prediction accuracy. The relative depth constraint provides greater weight on the depth axis and endows the model with the ability to perceive front-back relationships (positive outside the torso plane and negative otherwise), thereby alleviating the depth blur problem, which is an ability that pure 3D keypoint constraints do not possess. The results in Table~\ref{tab:aba} show that all losses contribute to improved performance.
 \begin{table}
\centering
\small
\begin{tabular}{lcccc}
\toprule[1.5pt]
\multirow{2}{*}{Method} & \multicolumn{2}{c}{3DPW-Test}  \\ 
 & MJE$\downarrow$  & V2V$\downarrow$  \\ \hline
HMMR~\cite{kanazawa2018end} & 116.5 & 72.6 \\ 
Doersch et al.~\cite{doersch2019sim2real} & - & 74.7 \\ 
Sun et al.~\cite{sun2019human} & - & 69.5 \\
TCMR~\cite{choi2021beyond} & 105.3 & 100.7 \\ 
VIBE~\cite{kocabas2020vibe} & 82.7 & 51.9 \\ 
MAED~\cite{wan2021encoder} & 79.1 & 45.7 \\ 
\hline
ReBaR & \textbf{69.1} & \textbf{41.8}  \\ \bottomrule[1.5pt]
\end{tabular}
\caption{\textbf{Evaluation on 3DPW-Test.} ReBaR achieves the best results without using the Euro filter."-" means no data provided.}
\label{tab:video}
\end{table}

\noindent \textbf{Compared To Video-based Methods}\\
We also compared our method to the state-of-the-art video-based methods. Table~\ref{tab:video} shows the results, which demonstrate that our method outperforms these methods by a significant margin even without additional temporal information from the video.
\begin{figure*}
  \centering
  \includegraphics[width=1\linewidth]{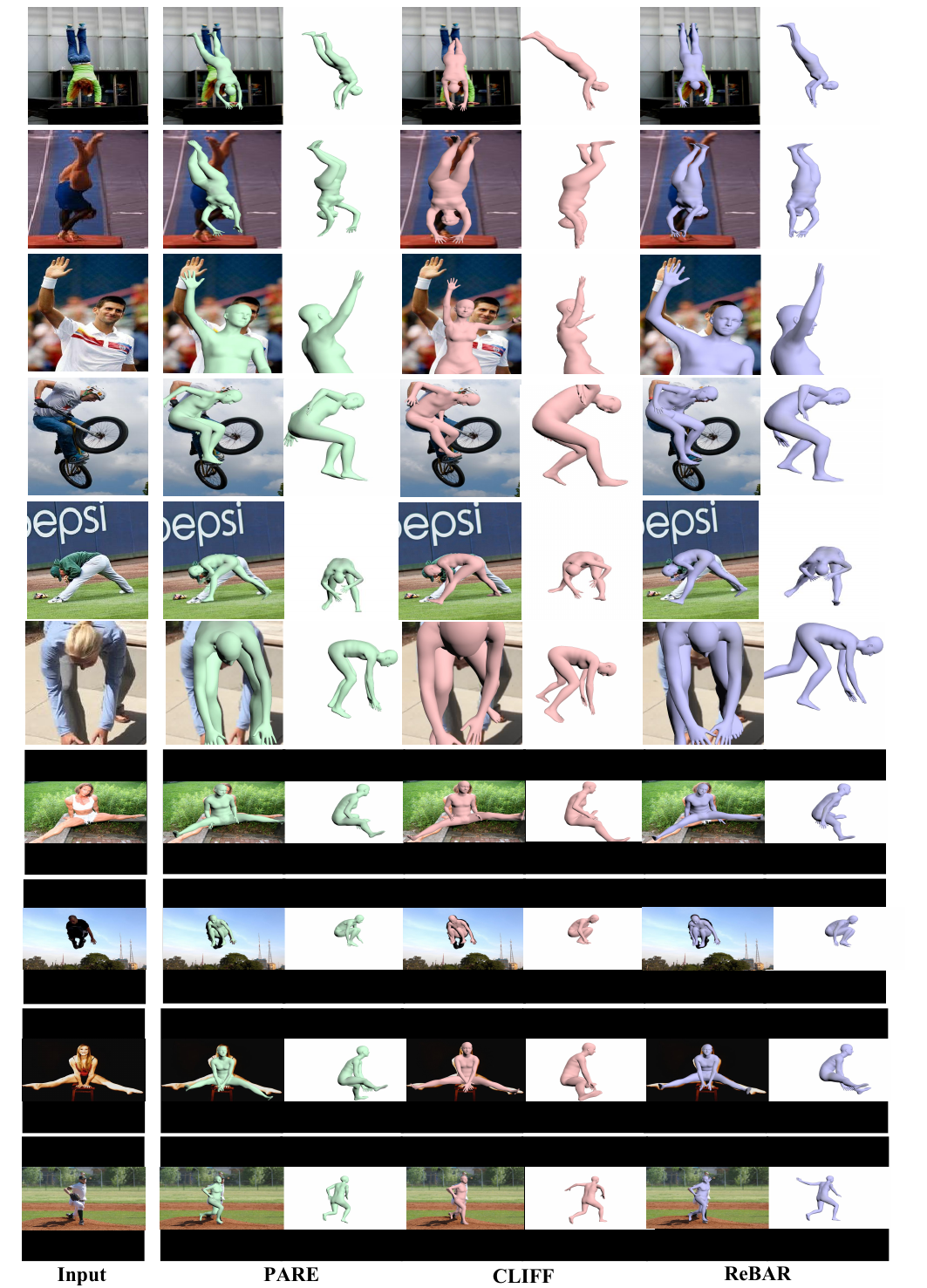}
  \caption{\noindent\textbf{Qualitative comparison of image.} From left to right: input image, PARE, CLIFF and ReBaR.}
  \label{fig:compare}
\end{figure*}
\section{More Qualitative Comparisons}
\label{sec:sec3}
In this section, we provide more qualitative comparison results. Visualize PARE~\cite{kocabas2021pare}, CLIFF~\cite{li2022cliff} and ReBaR results on images and challenging video files respectively for comparison. The image data includes two public datasets of 3DPW~\cite{von2018recovering} and LSPET~\cite{johnson2011learning}, which contain problems such as occlusion, direction blur, and depth blur. Video files are downloaded directly from the internet and include challenging action sequences such as yoga, hip-hop and more.

\noindent \textbf{Qualitative Comparison of Image.}\\
As shown in Figure~\ref{fig:compare}, our method outperforms PARE and CLIFF in almost all cases on the motion dataset LSPET. Especially under the problem of depth ambiguity and self-occlusion, thanks to the body-aware part features encoding, ReBaR can infer the spatial relationship between the occluded part and the body from the local visual cues around the part and the relevant global information of the whole body, Thereby improving the estimation accuracy. More interestingly, we found that PARE can barely determine the body orientation in some handstand situations, which we believe is due to PARE disconnecting the limbs and independently predicting the global rotation part. However, ReBaR avoids this problem nicely by using body reference conditions and limb dependencies.

\bibliographystyle{IEEEbib}
\bibliography{icme2023template}


\sloppy

\def\x{{\mathbf x}}
\def\L{{\cal L}}

\title{ExpGest: Expressive Full-Body Gesture Generation Using Diffusion Model and Hybrid Audio-Text Guidance\\ Supplementary Materials}
%
\name{Anonymous ICME submission}
\address{}

\maketitle

This supplementary material provides additional experimental results to reinforce the main paper. In Section~\ref{sec:sec1}, we present the visualization of similarity for the latent space semantic alignment module. Subsequently, in Section~\ref{sec:sec2}, we discuss some limitations of ExpGest and outline future plans. Finally, in Section~\ref{sec:sec3}, we demonstrate the application protocol based on Blender. In addition, we also present the use cases in the supplementary video. In its pioneering effort to seamlessly incorporate both motion and audio data to craft an expressive speaker, ExpGest presents remarkably expressive outcomes. We are confident that these compelling results will leave a lasting impact on researchers. Moreover, this endeavor simultaneously establishes the foundation for an innovative trajectory in the realm of audio-to-gesture applications.

\section{Visualization of the similarity matrix for SAM}
\label{sec:sec1}
\begin{figure}
  \begin{center}
  \vspace{-2em}
    \includegraphics[width=1\linewidth]{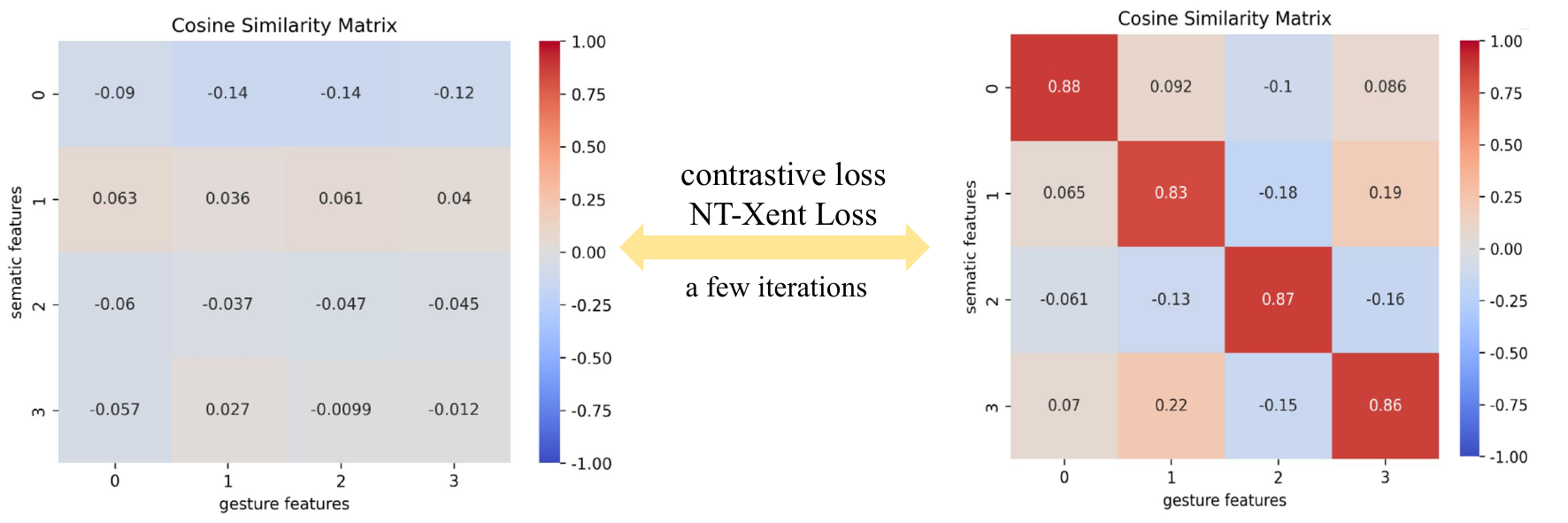}
    \vspace{-1em}
  \end{center}
  \vspace{-15pt}
  \caption{Illustration of training the gesture and transcription text latent space alignment module based on contrastive loss.}
  \vspace{-15pt}
  \label{fig:supp1}
  \end{figure}

\begin{figure}
  \begin{center}
  \vspace{-2em}
    \includegraphics[width=1\linewidth]{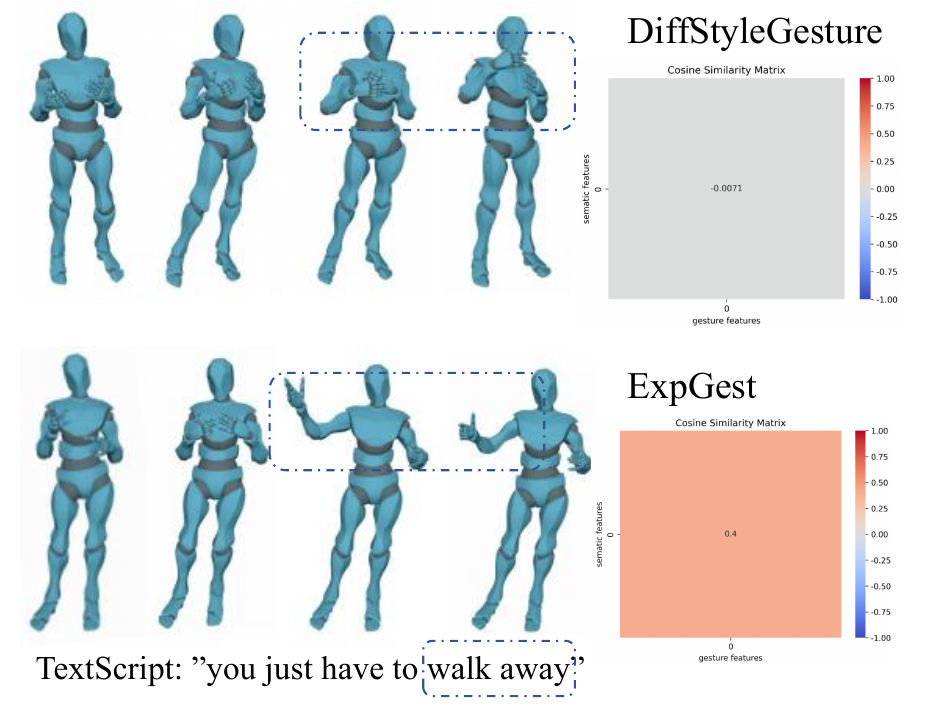}
    \vspace{-1em}
  \end{center}
  \vspace{-15pt}
  \caption{Displayed the semantic alignment degree of DiffStyleGesture and ExpGest on the test set.}
  \label{fig:supp2}
  \end{figure}

To further demonstrate the effectiveness of the latent space semantic-gesture alignment module, we retain the original training structure and freeze all parameter weights. SAM first employs the BERT~\cite{devlin2018bert} Tokenizer to encode each text segment and uses the hidden state of the first token ([CLS] tag) as the representation for that text segment, subsequently mapping it to a 256-dimensional space. Next, a pre-trained Gesture-VAE~\cite{kingma2013auto} is introduced to encode the real finger motion sequences into the latent space, with dimensions consistent with the text features. Finally, the cosine similarity between gesture features and semantic features is calculated (as shown in Equation 7 in the main text) and visualized as a heatmap. As illustrated in Figure~\ref{fig:supp1}, the encoded text features are highly aligned with the corresponding real gesture features, guiding the model to generate semantically related gesture sequences.\\
Compared to DiffuseStyleGesture without semantic alignment, the gestures generated by our method are more closely aligned with the speech content and can more accurately convey the talker's intent. We evaluated a subset of the BEAT test set, and as shown in Figure~\ref{fig:supp2}, ExpGest exhibits better semantic alignment scores (about 0.5), while DiffStyleGesture struggles to align (less than 0.1). Moreover, in a neutral context, DiffStyleGesture presents an almost stationary state.
\section{Limitations and Future Work}
\label{sec:sec2}
\begin{table}
\centering
\begin{adjustbox}{scale=0.85}
\begin{tabular}{lccccccccc}
\toprule[1.5pt]
\multirow{1}{*}{Method} &\multicolumn{1}{c}{$FGD (Feature)$} &\multicolumn{1}{c}{$EA$} &\multicolumn{1}{c}{$SA$} &\multicolumn{1}{c}{$Time(s)$}\\ \hline
DiffStyleGesture~\cite{yang2023diffusestylegesture}  &133.9 &0.60 &0.11 &0.102 \\ \hline
ExpGest (upper) &95.9 &0.57 &0.14 &0.082\\
ExpGest (full) &95.9 &0.57 &0.14 &0.157\\
ExpGest w/ SAM  &98.1 &0.62 &0.63 &0.181 \\
ExpGest w/ EG  &115.3 &0.89 &0.10  &0.45\\
ExpGest full model &\textbf{76.6} &\textbf{0.91} &\textbf{0.61}  &0.491\\
\hline
\end{tabular}
\end{adjustbox}
\vspace{-10pt}
\caption{Inference time statistics of ExpGest}
\vspace{-10pt}
\label{tab:times}
\end{table}
Despite the similarities between our method and MDM, which is based on the DDPM approach, speed and time cost remain significant challenges. We evaluated the inference speed of ExpGest, and as shown in Table~\ref{tab:times}, there is not much difference from DiffStyleGesture in the audio-only half-body mode. However, with the introduction of classifier guidance, each denoising step requires gradient backpropagation, leading to a substantial increase in time cost. Consequently, under the current circumstances, a real-time audio gesture generation system seems to be an impossible task. However, upon considering all generation methods, we observed an interesting phenomenon. GAN models can learn a relatively complex distribution, while the distribution from the 1000th step directly to the 0th step in Diffusion is so complex that GAN\cite{creswell2018generative} cannot learn it. Therefore, a compromise solution combining GAN and Diffusion for audio-to-gesture generation may make the generation results more efficient, meeting the requirements for real-time performance. Hence, in future work, we aim to explore a real-time Talker generator based on a Diffusion-GAN framework.
\section{Application Protocol for Blender}
\label{sec:sec3}
\begin{figure}
  \begin{center}
  \vspace{-2em}
    \includegraphics[width=1\linewidth]{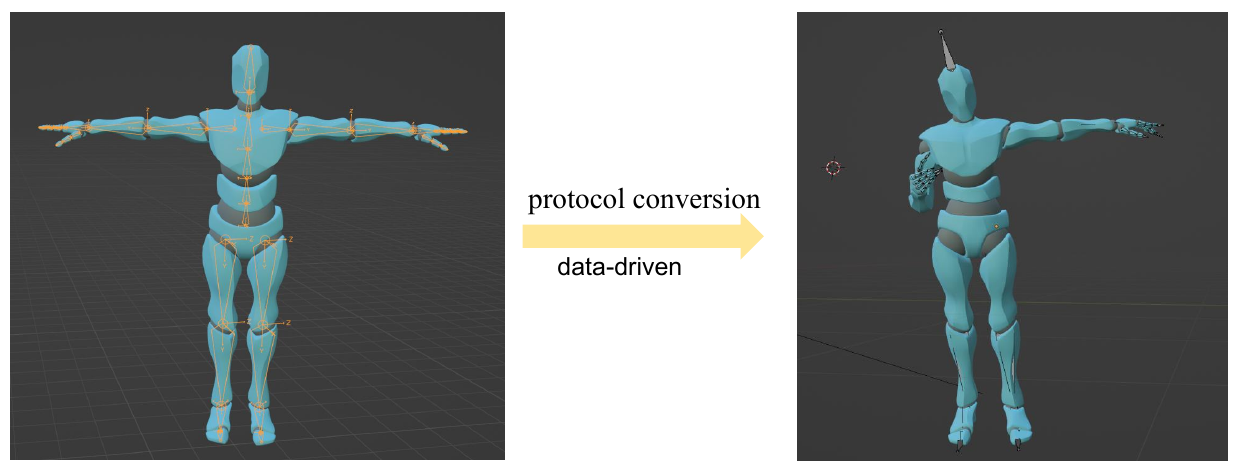}
    \vspace{-1em}
  \end{center}
  \caption{Illustration of the coordinate protocol alignment and data-driven process for FBX character models}
  \label{fig:supp3}
  \end{figure}

In this section, we introduce the application tool, a virtual human~\cite{han2023hutumotion} driving script created with Blender, designed to use the results of ExpGest to drive any downloaded or created FBX character model. As illustrated in Figure~\ref{fig:supp3}, we first import the character model into Blender and enter pose mode. We select the bones corresponding to the SMPLX~\cite{loper2015smpl} joints and uniformly convert them to a coordinate protocol with the X-axis pointing right, the Y-axis pointing up, and the Z-axis pointing forward. Subsequently, we adjust the character model with the converted protocol to the initial T-Pose posture. Finally, by invoking Blender's bpy script, we match the joint names of the character model with the SMPLX joint indices and convert the original axis angle to quaternion to complete the character model driving. The entire repository, including models and applications, will be made available after the conference results are announced.

\bibliographystyle{IEEEbib}
\small
\bibliography{icme2023template}